\documentclass{article}

\usepackage{PRIMEarxiv}

\usepackage[utf8]{inputenc} 
\usepackage[T1]{fontenc}    
\usepackage{hyperref}       
\usepackage{url}            
\usepackage{booktabs}       
\usepackage{amsfonts}       
\usepackage{nicefrac}       
\usepackage{microtype}      
\usepackage{fancyhdr}       
\usepackage{graphicx}       
\graphicspath{{../figures/}} 

\usepackage{makecell}
\usepackage{amsmath}
\usepackage{multirow}
\usepackage{enumitem}
\usepackage{array}
\usepackage{appendix}
\usepackage{colortbl}
\usepackage{times}
\usepackage{latexsym}
\usepackage{pifont}
\newcommand{\cmark}{\ding{51}}%
\newcommand{\xmark}{\ding{55}}%
\usepackage{siunitx}
\usepackage[dvipsnames]{xcolor}
\definecolor{green}{rgb}{0.0, 0.6, 0.0}
\definecolor{brightgreen}{rgb}{0.0, 0.9, 0.0}
\usepackage{natbib}
\usepackage{caption}

\title{Test-Time Scaling Strategies for Generative Retrieval in Multimodal Conversational Recommendations}

\author{
  \textbf{Hung-Chun Hsu}\textsuperscript{1}\thanks{Main author. Email: r10946017@citi.sinica.edu.tw}, \textbf{Yuan-Ching Kuo}\textsuperscript{1}, \textbf{Chao-Han Huck Yang}\textsuperscript{2}, 
  \textbf{Szu-Wei Fu}\textsuperscript{2}, \\ \textbf{Hanrong Ye}\textsuperscript{2}, \textbf{Hongxu Yin}\textsuperscript{2}, 
  \textbf{Yu-Chiang Frank Wang}\textsuperscript{2}, \textbf{Ming-Feng Tsai}\textsuperscript{3}, \textbf{Chuan-Ju Wang}\textsuperscript{1}\thanks{Corresponding author. Email: cjwang@citi.sinica.edu.tw} \\
  \\
  \textsuperscript{1}Research Center for Information Technology Innovation, Academia Sinica, Taiwan\\
  \textsuperscript{2}NVIDIA\\
  \textsuperscript{3}Department of Computer Science, National Chengchi University, Taiwan\\
}

\begin{document}
\maketitle

\begin{abstract}
The rapid evolution of e-commerce has exposed the limitations of traditional product retrieval systems in managing complex, multi-turn user interactions.
Recent advances in multimodal generative retrieval---particularly those leveraging multimodal large language models (MLLMs) as retrievers---have shown promise.
However, most existing methods are tailored to single-turn scenarios and struggle to model the evolving intent and iterative nature of multi-turn dialogues when applied naively.
Concurrently, test-time scaling has emerged as a powerful paradigm for improving large language model (LLM) performance through iterative inference-time refinement.
Yet, its effectiveness typically relies on two conditions: (1) a well-defined problem space (e.g., mathematical reasoning), and (2) the model's ability to self-correct---conditions that are rarely met in conversational product search. 
In this setting, user queries are often ambiguous and evolving, and MLLMs alone have difficulty grounding responses in a fixed product corpus.
Motivated by these challenges, we propose a novel framework that introduces test-time scaling into conversational multimodal product retrieval.
Our approach builds on a generative retriever, further augmented with a test-time reranking (TTR) mechanism that improves retrieval accuracy and better aligns results with evolving user intent throughout the dialogue.
Experiments across multiple benchmarks show consistent improvements, with average gains of 14.5 points in MRR and 10.6 points in nDCG@1.
\end{abstract}


\section{Introduction}

\begin{figure}[t]
    \centering
    \includegraphics[width=0.55\columnwidth]{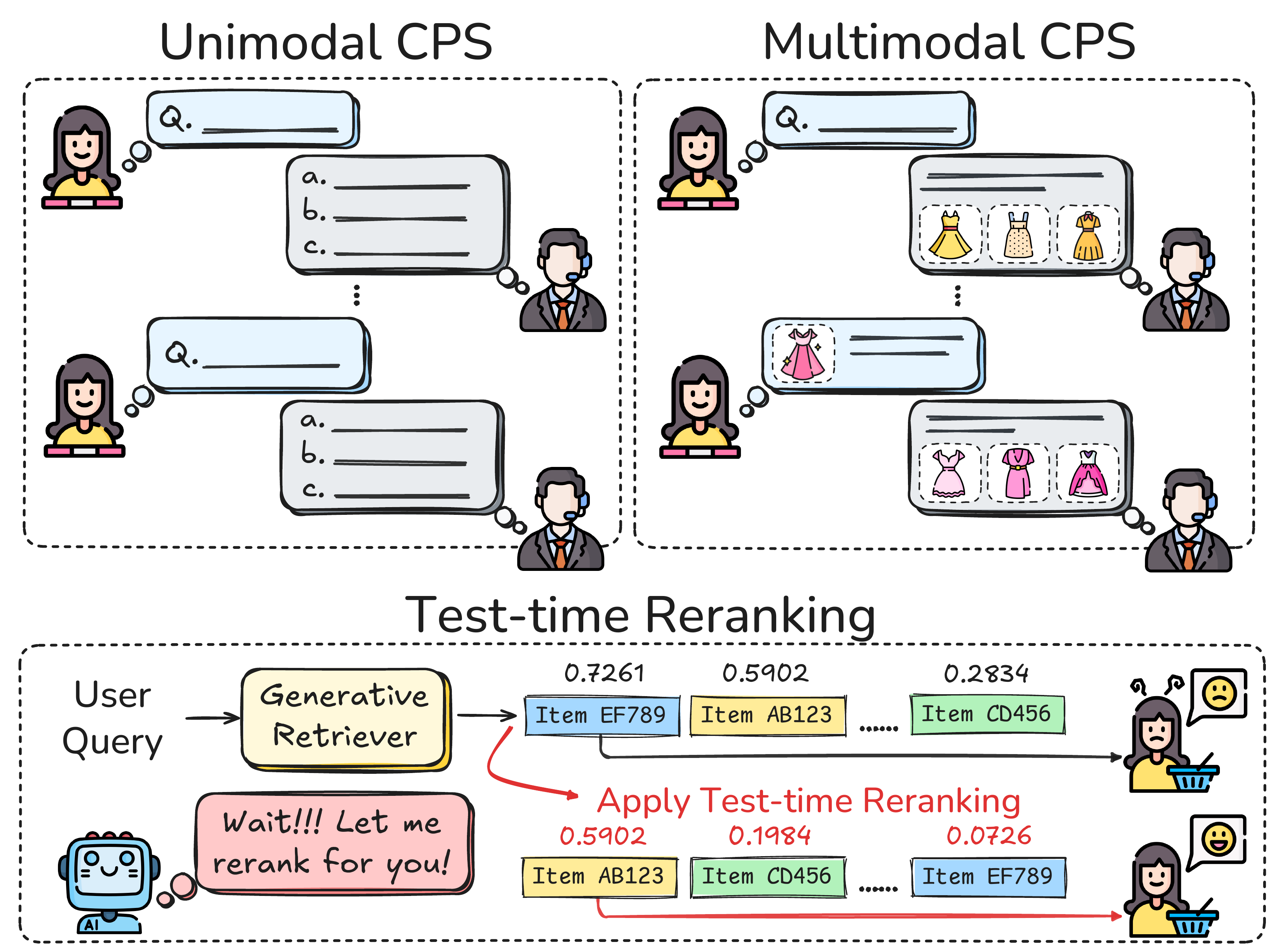}
    \vspace{0.3cm}
    \caption{\textbf{Top: Unimodal and multimodal conversational product search (CPS).} Multimodal CPS incorporates user-provided reference images to more effectively retrieve the desired product. \textbf{Bottom: Overview of the proposed test-time reranking (TTR) mechanism.} As the dialogue progresses, product scores produced by a pre-trained LLM-based generative retriever are dynamically adjusted at inference time to better align with the evolving user intent.
}
    \label{fig:front_page_diagram}
\end{figure}

{Conversational search} marks a significant departure from traditional single-turn information retrieval by enabling users to satisfy complex information needs through multi-turn dialogues~\cite{10.1145/3020165.3020183}.
A central challenge in this paradigm is interpreting user queries in the context of preceding dialogue history.
To address this, various query reformulation techniques---such as query expansion~\citep{wang-etal-2023-query2doc, ma-etal-2023-query} and query rewriting~\cite{10.1145/3397271.3401323, mo-etal-2023-convgqr}---have been developed to bridge user intent across turns.

In the e-commerce domain, conversational product search (CPS) extends these principles to transform how users engage with product search systems~\cite{mo2024surveyconversationalsearch}.
Unlike traditional keyword-based systems that mostly treat each query independently and often struggle with evolving or ambiguous user intent~\cite{mo2024surveyconversationalsearch}, CPS leverages natural language dialogues to iteratively surface fine-grained user preferences~\cite{10.1145/3269206.3271776, 10.1145/3571884.3604318}.
By preserving conversational context and enabling proactive intent clarification, CPS systems can better constrain the search space and align retrieval results with user needs. 
This context-aware, interactive process ultimately enhances product discovery, improves recommendation accuracy, fosters a richer, more engaging user experience~\cite{mo2024surveyconversationalsearch, 10.1145/3589334.3645483}.

Building on the foundations of CPS, recent studies have begun to explore how the generative capabilities of large language models (LLMs) can enhance retrieval effectiveness.
Generative retrieval~\cite{decao2021autoregressive, bevilacqua2022autoregressive, 10.5555/3600270.3601857}---leveraging the autoregressive nature of LLMs to produce relevant document or item identifiers---has shown promise in various domains, including e-commerce search~\cite{li2024generativeretrievalpreferenceoptimization, lin2025efficient} and cross-modal retrieval~\cite{li-etal-2024-generative, DBLP:conf/iclr/LinLSLCP25}.
In particular, multimodal conversational retrieval is gaining traction, as users increasingly express preferences through diverse modalities such as text, images, videos, or speech.
The upper panel of Figure~\ref{fig:front_page_diagram} contrasts unimodal CPS with multimodal CPS, the latter incorporating user-provided reference images to retrieve more relevant products.
Despite the conceptual alignment between generative retrieval and the goals of CPS, this intersection remains underexplored---especially in multi-turn product search and recommendation scenarios. 
A key barrier is the scarcity of publicly available datasets that support multimodal, multi-turn retrieval tasks, thereby hindering systematic research and benchmarking.

On the other hand, despite its potential, generative retrieval remains constrained in practice---particularly in multimodal conversational settings---due to the fixed computational budget typically allocated at inference.  
This constraint motivates the use of {test-time scaling (TTS)}, or test-time compute (TTC), which aims to improve the performance of pre-trained models by allocating additional computation during inference~\cite{snell2024scalingllmtesttimecompute, muennighoff2025s1simpletesttimescaling, liu20251bllmsurpass405b}.
Given that generative retrievers are usually built on LLMs, TTS techniques serve as a natural extension to enhance their performance.
Unlike approaches that scale model size or training data, TTS focuses on enabling more sophisticated reasoning during inference, thereby improving output quality on complex tasks.

To advance scalable multimodal conversational retrieval, we propose a novel framework that integrates a multimodal LLM-based generative retriever with an effective \textbf{test-time reranking} (TTR) mechanism.
This design bridges CPS with recent advances in TTS~\cite{mo2024surveyconversationalsearch}, enabling more accurate retrieval in multi-turn dialogue settings.
As illustrated at the bottom of Figure~\ref{fig:front_page_diagram}, the TTR component adjusts the scores of candidate items from the generative retriever, refining the final output.
To support training and evaluation, we curate and refine two existing datasets---Multi-turn Fashion IQ~\cite{10.1145/3404835.3462881} and Multimodal Dialogue (MMD) \cite{10.5555/3504035.3504121}---and additionally incorporate MUSE ~\cite{wang-etal-2025-muse}, a synthetic multimodal conversational recommendation dataset.
Our contributions are:

\begin{itemize}[leftmargin=*]
\setlength{\parskip}{0pt}
  \setlength{\itemsep}{0pt plus 1pt}
    \item We introduce a novel framework that integrates generative retrieval with multimodal, multi-turn dialogue, addressing an underexplored area in conversational product search.
    \item We propose \textbf{test-time reranking} (TTR), a test-time scaling mechanism for product reranking, enabling dynamic refinement during inference and improving retrieval quality in complex, multi-turn multimodal scenarios.
    \item We curate and release improved multi-turn multimodal product retrieval datasets---refining existing benchmarks and incorporating a synthetic corpus---to support research and evaluation in this emerging domain.
    \item 
    We conduct extensive experiments on three benchmarks, demonstrating that our framework consistently outperforms strong unimodal and multimodal baselines.
\end{itemize}

\section{Related Works}
\subsection{Conversational Product Search}
A key challenge in conversational retrieval is leveraging complex, evolving dialogue context to retrieve relevant results~\cite{mo2024surveyconversationalsearch}.
One common approach reformulates conversational queries into standalone inputs suitable for standard retrieval models.
For example, ConvGQR utilizes pre-trained language models (PLMs) for query rewriting~\cite{mo-etal-2023-convgqr}, while CONQRR employs reinforcement learning to rewrite queries specifically for retrieving passages in conversational question answering (CQA)~\cite{wu-etal-2022-conqrr}.
However, such reformulations may misalign downstream retrieval objectives and struggle with lengthy or ambiguous conversations.

Conversational dense retrieval offers an alternative by directly encoding dialogue history.
ChatRetriever~\cite{mao-etal-2024-chatretriever} combines instruction tuning with contrastive learning to adapt large language models (LLMs) for this task, improving the modeling of multi-turn interactions.
Applying these techniques to multimodal conversational product search introduces additional complexity: handling multimodal queries and corpora.
Real-world product queries often involve not only text but also images and structured attributes---capabilities that text-only models typically lack.

Compounding this issue is the lack of suitable multimodal conversational retrieval datasets.
Existing resources are limited: MG-ShopDial~\cite{Bernard_2023} is text-only; Melon~\cite{10.1145/3589334.3645483} focuses on clarifying questions rather than retrieval; and Multiturn Fashion Retrieval (MFR)~\cite{10.1145/3471158.3472257, 10.1145/3543507.3583420} does not exhibit true conversational coherence. 
Also, the Multimodal Dialogue (MMD) dataset~\cite{10.5555/3504035.3504121} is no longer maintained and contains numerous broken data links.
Synthetic alternatives like MUSE \cite{wang-etal-2025-muse} offer promise but cannot fully replace real-world data.
While MFR, MMD, and MUSE remain valuable testbeds for current research, these limitations highlight the urgent need for more comprehensive multimodal datasets to advance the field.

\subsection{Generative Document Retrieval}

Conventional information retrieval (IR) methods rely on matching user queries against indexed document collections using techniques such as sparse retrieval (e.g., BM25~\cite{10.1561/1500000019}) or dense bi-encoder architectures (e.g., DPR~\cite{karpukhin-etal-2020-dense}, ANCE~\cite{xiong2021approximate}) based on vector similarity. In contrast, generative retrieval (GR) marks a significant paradigm shift by leveraging LLMs to produce relevant document identifiers (DocIDs) using a sequence-to-sequence framework~\cite{10.1145/3722552}.
This approach enables a novel mechanism for information access, using generation strategies like constrained beam search to restrict outputs to valid DocIDs, which are then ranked by generation likelihood.

DocIDs in GR systems can be structured (e.g., via recursive document clustering techniques \cite{10.5555/3600270.3601857}) or semantic, often referred to as Semantic IDs (SIDs), using key text fragments or image captions~\cite{li-etal-2024-generative, li-etal-2023-multiview}.
GR has shown promise in diverse applications, including cross-modal retrieval \citep{10.1609/aaai.v38i17.29837, li2025semcoresemanticenhancedgenerativecrossmodal} and recommender systems \cite{rajput2023recommender, wang2024generativerecommendationnextgenerationrecommender}.
Ranking-based objectives can further improve performance~\cite{10.1609/aaai.v38i8.28717, jin2025llm}.
To address scalability challenges, GR systems can integrate efficient data structures such as the FM-index~\cite{892127}, which enables fast and compact retrieval within large text corpora~\cite{bevilacqua2022autoregressive}.

\subsection{Test-time Scaling}

Test-time scaling (TTS) is an emerging paradigm  that enhances LLMs performance by allocating additional computation during inference~\cite{snell2024scalingllmtesttimecompute, muennighoff2025s1simpletesttimescaling}. Rather than relying solely on model or data scale, TTS improves output quality by generating multiple candidate responses and selecting the most promising one.
Selection is typically supported by external verification model~\cite{liang2024improvingllmreasoningscaling, lee2025reviselearningrefinetesttime, wang-etal-2024-math}, with reinforcement learning further improving model self-correction~\cite{kumar2024traininglanguagemodelsselfcorrect, mendes2025languagemodelsselfimprovestatevalue}.

TTS strategies generally fall into two categories: (1) expanding chain-of-thought (CoT) reasoning via longer paths (e.g., tree-of-thoughts) \citep{10.5555/3600270.3602070, bi2025forestofthought}, and (2) multi-sample generation methods such as best-of-N (BoN) sampling \citep{lightman2024lets, wang2025samplingefficienttesttimescalingselfestimating} or Beam Search \citep{snell2024scalingllmtesttimecompute, 10.5555/3692070.3694110}.

In retrieval context, TCR~\cite{li2025testtime} addresses query shift in image-text retrieval but is limited to single-turn tasks. 
Similarly, Rank1~\cite{weller2025rank1testtimecomputereranking} uses TTS to rerank documents by distilling reasoning traces, 
showing gains on MS MARCO~\cite{DBLP:conf/nips/NguyenRSGTMD16} and BRIGHT~\cite{su2025brightrealisticchallengingbenchmark}, yet it remains constrained to single-turn unimodal ranking.
These works highlight TTS's potential in retrieval but also expose a gap: a lack of approaches for multi-turn conversational retrieval, especially in multimodal contexts.

\begin{figure*}[h] 
    \centering
    \includegraphics[width=\textwidth]{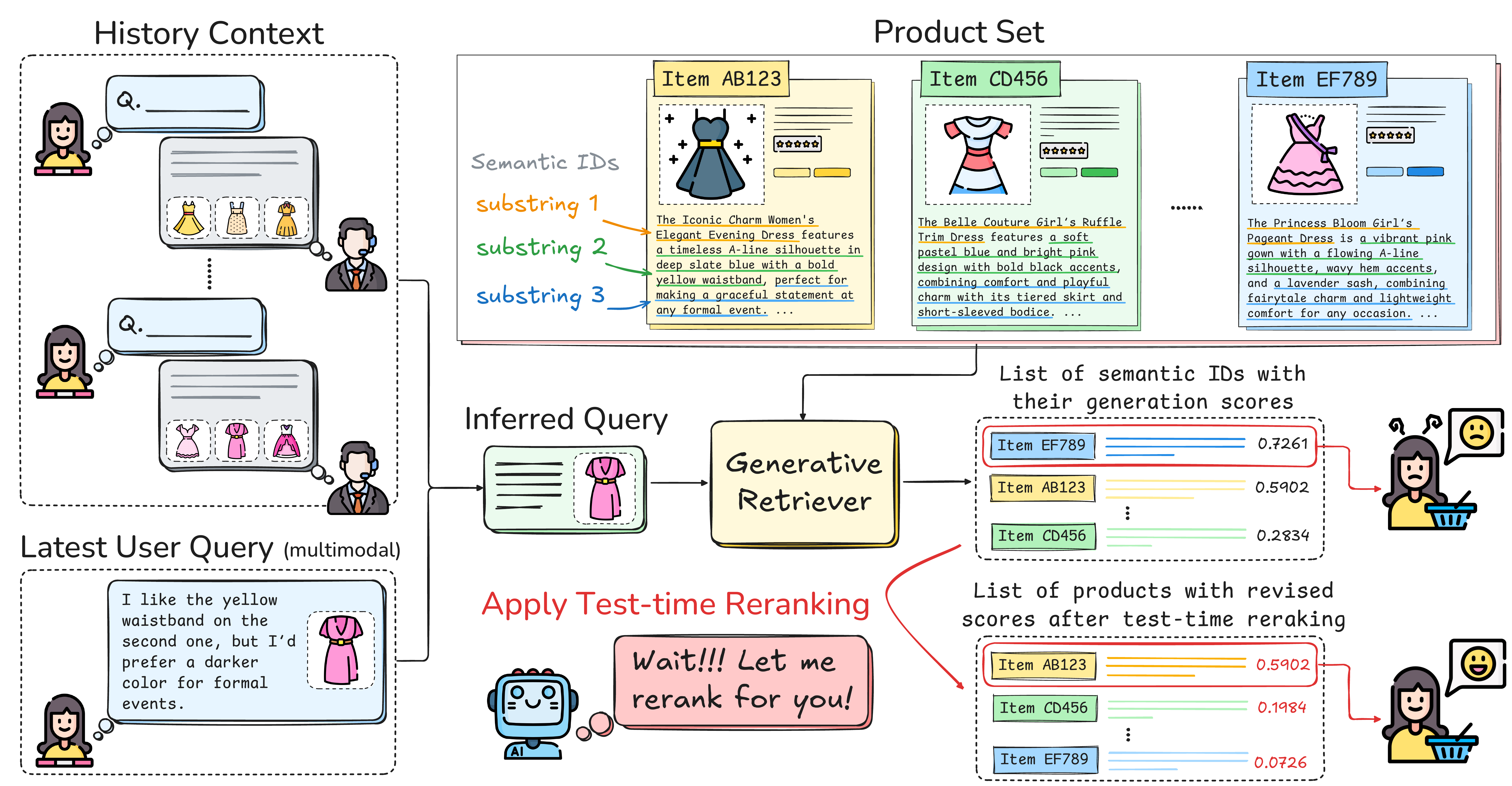}
    \vspace{0.3cm}
    \caption{Overview of the proposed three-stage pipeline for conversational multimodal product search enhanced by test-time scaling: (i) \textbf{User Intent Inference}: A multimodal large language model (e.g., GPT-4o-mini) infers the user's current intent based on the dialog context. (ii) \textbf{Semantic ID-based Generative Retrieval}: The inferred intent, combined with a user-provided reference image, is used by a generative retriever to produce semantic identifiers of relevant products via beam search over an FM-index-constrained output space. (iii) \textbf{Test-time Reranking}: The retreived products are further refined using our proposed test-time reranking mechanism that adjusts their scores based on alignment with the inferred user intent.}
    \label{fig:framework_overview}
\end{figure*}

\section{Methodology}
In this section, we introduce our framework for test-time conversational multimodal product search. 
We begin by defining the task and then describe the framework's key components and training strategy.
Finally, we explain how it supports scalable computation during inference.

\subsection{Task Definition}
Conversational multimodal product search requires a retrieval framework capable of iteratively refining retrieval results based on the user's evolving requests---such as queries or feedback---across multiple dialogue turns.
Formally, let $X = \{x_i\}_{i=1}^N$ be a corpus of $N$ multimodal products, where each product $x_i = (x_i^{\rm text}, x_i^{\rm img})$ contains both textual and visual modalities.
To support multimodal interaction, we allow user utterances and system responses to vary in modality, with each turn associated with a modality set $\mathcal{M} \subseteq \{\mathrm{text}, \mathrm{img}\}$.
At dialogue turn~$t$, the system aims to retrieve a target product by leveraging the dialogue history $\{q_1^{\scriptscriptstyle \mathcal{M}}, p_1^{\scriptscriptstyle \mathcal{M}}, \ldots, q_{t-1}^{\scriptscriptstyle \mathcal{M}}, p_{t-1}^{\scriptscriptstyle \mathcal{M}}\}$ (denoted $d_{<t}^{\scriptscriptstyle \mathcal{M}}$), along with the current user query $q_t^{\scriptscriptstyle \mathcal{M}}$. 
For example, $q_t^{\scriptscriptstyle \mathcal{M}} = (q_t^{\rm text}, q_t^{\rm img})$ comprises both textual and visual content.

A query reformulator $f_Q$ transforms the current dialogue context $(d_{<t}^{\scriptscriptstyle \mathcal{M}}, q_t^{\scriptscriptstyle \mathcal{M}})$ into an inferred query.
A similarity function $s(\cdot)$ is then used to score all candidate products in the corpus $X$ against this representation.
The top-$K$ scoring products are returned as the result at turn $t$, denoted $\mathcal{P}_{t}^K$:

\begin{equation*} \label{eq:multi-turn-retrieval}
\mathcal{P}_{t}^K = \underset{i \in \{1,2,...,N\}}{\text{top-}K} \, s\left(f_Q\big(d_{<t}^{\scriptscriptstyle \mathcal{M}}, q_t^{\scriptscriptstyle \mathcal{M}} \big), x_i\right).
\end{equation*}

The goal of the conversational multimodal product search is to retrieve the user's intended product at each turn, despite challenges such as underspecified queries, ambiguous references, and evolving preferences throughout the dialogue.

\subsection{Multimodal Conversational Product Retriever}
Our multimodal conversational product retriever consists of three key modules. 
First, an intent understanding module utilizes a large language model (LLM) to interpret the user's current intent at each dialogue turn.
Depending on the setting, this can be powered by either proprietary or open-source LLMs. 
In multimodal contexts, we adopt multimodal LLMs capable of processing inputs that include text alongside visual, video, or audio signals.
Second, a retrieval module uses the interpreted intent to generate an initial set of candidate product identifiers from the product corpus.
Finally, a test-time reranking (TTR) module refines these candidates by reranking them based on their semantic relevance to the user's query.
This reranker incorporates a test-time evaluation mechanism for improved precision. 
The overall architecture is illustrated in Figure~\ref{fig:framework_overview}.

\subsubsection{User Intent Inference}\label{sec:inferred}
Given the dialogue history $d_{<t}^{\scriptscriptstyle \mathcal{M}}$ and the current user query $q_t^{\scriptscriptstyle \mathcal{M}}$, an LLM is utilized to infer the user's intent at turn $t$, producing a revised query representation $\widehat{q}^{\scriptscriptstyle \mathcal{M}}_t$.
Formally, we generate a textual reformulation as
$\widehat{q_t}^{\rm text} = \text{LLM}(d_{<t}^{\scriptscriptstyle \mathcal{M}}, q_t^{\scriptscriptstyle \mathcal{M}})$, which is then combined with the original visual input (if present), yielding the multimodal inferred query: $\widehat{q}_t^{\scriptscriptstyle \mathcal{M}} = (\widehat{q_t}^{\rm text}, q_t^{\rm img})$.
This inferred query is subsequently passed to the generative retriever $f_{\rm retriever}$, which we describe in the next subsection.

\subsubsection{Semantic ID-based Generative Retriever}\label{sec:generative_retriever}

We employ a generative model as our retriever, which outputs semantic identifiers (SIDs) to represent each product. 
These SIDs are constructed from rich product information such as attribute values (e.g., style, shape, price) and image captions. 
Following the approach~\cite{li-etal-2023-multiview}, we adopt substrings of product descriptions as SIDs. 
Each product may be associated with multiple SIDs, each capturing different aspects of it (e.g., $\text{product}\ x_1: \text{SID}_a$, $\text{product}\ x_1: \text{SID}_b$).
The generative retriever, $f_{\text{retriever}}$, is a language model fine-tuned on dialogue data to generate relevant SIDs for a specific product under output constraints in response to the inferred query.

During inference, given an inferred query $\widehat{q}^{\scriptscriptstyle \mathcal{M}}$ and a candidate product $x$, the generative retriever $f_{\text{retriever}}$ produces the $b$ most semantically relevant SID(s) for product $x$ with respect to the inferred query, denoted as:

\begin{equation*}
C_x = \{c_1, \dots, c_j, \dots, c_b\} = f_{\rm retriever}(\widehat{q}^{\scriptscriptstyle \mathcal{M}} \mid x).
\end{equation*}

\begin{figure}[h]
    \centering
    \includegraphics[width=0.6\columnwidth]{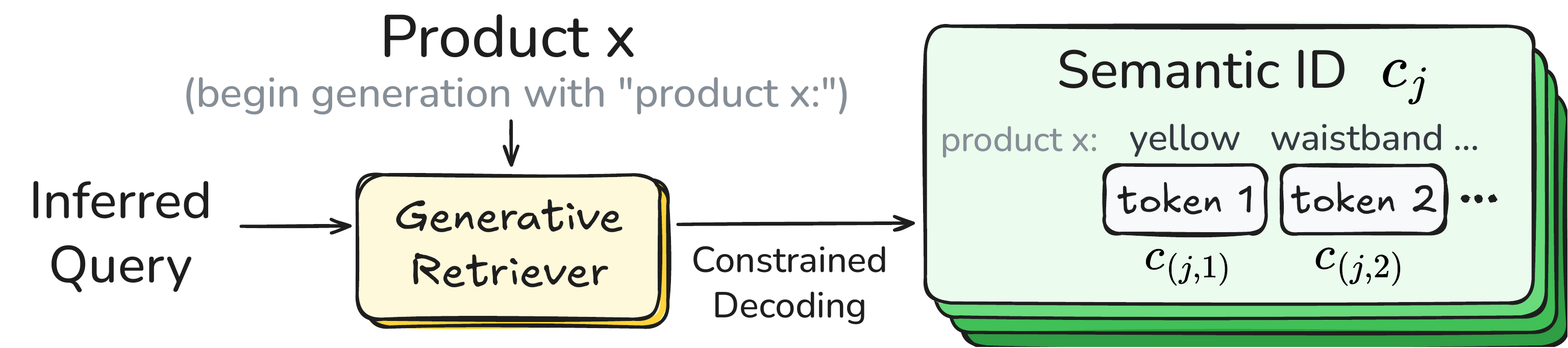}
    \vspace{0.3cm}
    \caption{Each SID $c_j \in C_x$ of product $x$ is represented as a sequence of tokens: $c_j = \{ c_{(j, 1)}, c_{(j, 2)}, \dots, c_{(j, |c_j|)} \}$. The generation probability score of $c_j$ is defined as the sum of the log probabilities of its constituent tokens: $\text{RM}(c_j) = \log(c_j) \stackrel{\text{def}}{=} \sum\log(c_{(j, :)})$.}
    \label{fig:example_generation_probability}
\end{figure}

Each SID $c_j \in C_x$ is accompanied by a probability score produced by the retriever during generation. 
In the context of test-time scaling, the retriever's score can be viewed as a reward signal that reflects its confidence in generating a SID~$c_j$ associated with product $x$, conditioned on the inferred query $\widehat{q}^{\scriptscriptstyle \mathcal{M}}$.
For simplicity, we denote this reward as $\text{RM}(c_j) = \log(c_j)$, where $\log(c_j)$ represents the sum of the log probabilities of the tokens comprising $c_j$.
Figure~\ref{fig:example_generation_probability} illustrates the computation of a SID's probability.

The final relevance score between a product $x$ and the query $\widehat{q}^{\scriptscriptstyle \mathcal{M}}$ is computed based on the scores of the associated SIDs.
While various aggregation strategies are possible—such as averaging or taking the maximum of the scores—we adopt the maximum generation score as the similarity measure between the product and the query:

\begin{equation*}
s(\widehat{q}^{\scriptscriptstyle \mathcal{M}}, x) = \arg\max_{c_j \in C_x} ~\text{RM}(c_j).
\end{equation*}

\subsection{Test-time Reranking}\label{sec:test_time_reranking}
To enhance generative retriever performance, we propose \textbf{test-time reranking} (TTR), a novel method inspired by recent advances in adapting test-time scaling to LLMs.
To the best of our knowledge, TTR is the first approach to apply test-time scaling in conversational product search, enabled by the underlying generative retrieval framework.
Specifically, TTR augments the original reward model by incorporating a test-time evaluator, yielding a revised scoring function that is subsequently used to rerank candidate items:

\begin{equation}
\text{RM}_{\rm TTR}(c_j) = \sigma(\log(c_j)) \times w_{\text{Eval}}(c_j \mid \widehat{q}^{\scriptscriptstyle \mathcal{M}}),
\label{eq:test-time product reranking}
\end{equation}

where $\sigma$ denotes a min-max normalization function applied to the log-probability scores, scaling them to the range $[0, 1]$ for comparability. 
In parallel, $w_{\text{Eval}}(c_j \mid \widehat{q}^{\scriptscriptstyle \mathcal{M}})$ represents the evaluator's confidence in how well the candidate semantic ID (SID) $c_j$ aligns with the inferred query $\widehat{q}^{\scriptscriptstyle \mathcal{M}}$, expressed as a scalar confidence score in the range $[0, 1]$.

\paragraph{Efficiency Analysis of Test-time Reranking (TTR)}
Efficiency and computational overhead are key concerns when applying TTR.
We analyze the computational cost introduced by the LLM-based evaluator, which constitutes the primary source of additional inference-time overhead.
The overall complexity of TTR is dominated by the number of candidate products $A$ and the number of retrieved (as well as generated) SIDs per product $B$, resulting in a total complexity of $O(AB)$.

To generate candidate SIDs, we adopt beam search for its favorable trade-off between computational efficiency and the ability to generate high-probability sequences. The proposed evaluation process is highly parallelizable and can be efficiently executed via multithreading, demonstrating the practical viability of our test-time reranking approach.

\section{Multimodal CPS Dataset}
Publicly available datasets for multi-turn multimodal retrieval are not only scarce but also suffer from notable limitations, making robust evaluation in this domain particularly challenging. 
Alongside our proposed framework for conversational product search (CPS), we contribute curated datasets specifically tailored to the demands of complex, real-world multimodal CPS scenarios.

The Multiturn Fashion Retrieval (MFR) dataset~\cite{10.1145/3404835.3462881}, derived from FashionIQ~\cite{Wu_2021_CVPR}, lacks semantic richness.
Each product is described by an image and a limited set attributes (e.g., style, fabric, and shape), without detailed product descriptions and meaningful captions.
This textual sparsity hinders fine-grained product matching and retrieval~\cite{zhao-etal-2024-unifashion}. 
To address this, we enhanced MFR with image captions released by UniFashion~\cite{zhao-etal-2024-unifashion}, generated using LLaVA-NeXT~\cite{liu2024llavanext} for FashionIQ images. We refer to this curated version as $\text{MFR}_{\text{crt}}$ in the rest of the paper.

Similarly, while the Multimodal Dialogs (MMD) dataset~\cite{10.5555/3504035.3504121} contains over 150,000 expert-curated conversational sessions in the fashion retail domain, many of its product URLs are now invalid.
To address this, we reconstructed a usable subset of MMD by filtering for valid product links, referred to as $\text{MMD}_{\text{flt}}$ in our experiments.
We further curated this subset by removing dialogues that were unsuitable for multimodal product retrieval---such as those in which the user never selected or purchased a product.

Beyond real-world datasets, we incorporate a synthetic corpus to examine LLM-generated dialogues in retrieval tasks. Recent work has proposed frameworks that simulate multimodal recommendation dialogues by modeling user-agent interactions with LLMs. We follow MUSE~\cite{wang-etal-2025-muse}, which constructs synthetic dialogues using simulated user profiles and a rich catalog of fashion products (e.g., clothing, shoes, and jewelry), and generate our own MUSE dataset to assess the generalizability of retrieval models across real-world and synthetic settings.

After curation, all datasets are standardized into a unified format, where each dialogue is represented as  
$\{q_1^{\scriptscriptstyle \mathcal{M}}, p_1^{\scriptscriptstyle \mathcal{M}}, \ldots, q_{t}^{\scriptscriptstyle \mathcal{M}}, p_{t}^{\scriptscriptstyle \mathcal{M}}, \ldots\}$.  
Each user query $q_t^{\scriptscriptstyle \mathcal{M}}$ may include both a textual query $q_t^{\rm text}$ and a reference image $q_t^{\rm img}$ explicitly mentioned by the user--- typically one of the product images recommended by the system in previous turns.  
Since the FM-index used in our retrieval pipeline supports only textual objects, each response $p$ is represented using the product's image caption and associated textual metadata (e.g., product ID and description).  
A summary of dialogue statistics is provided in Table~\ref{tab:dataset_statistics}. 

\begin{table}[htbp]
\centering
\begin{tabular}{@{}l r r r r r@{}}
\toprule
\thead{Dataset} & \thead[r]{\# Dialogues} & \thead[r]{Train} & \thead[r]{Valid} & \thead[r]{Test} & \thead[r]{\# Products} \\
\midrule
\texttt{$\text{MFR}_{\text{crt}}$} & 2,350 & 1,809 & 441 & 100 & 15,269 \\
\texttt{$\text{MMD}_{\text{flt}}$} & 10,000 & 9,400 & 500 & 100 & 49,164 \\
\texttt{MUSE} & 7,000 & 6,500 & 400 & 100 & 94,209 \\
\bottomrule
\end{tabular}
\vspace{0.3cm}
\caption{Dataset statistics.}
\label{tab:dataset_statistics}
\end{table}

\section{Experiments}

\subsection{Baselines}\label{sec:baseline}

To evaluate our proposed framework for multimodal conversational product search (CPS), 
we assess its impact on retrieval performance across several widely used retrievers.
A distinguishing feature of our framework is that its support for multimodal queries, allowing it to incorporate input such as user-referenced images from earlier dialogue turns.
In contrast, the baseline unimodal retrievers are restricted to text-only inputs.
These include dense retriever  
T5-Large~\cite{10.5555/3455716.3455856} and LLM-based retriever Qwen2.5-7B~\cite{bai2025qwen25vltechnicalreport}. 
For the multimodal setting, we evaluate VLT5-Large~\cite{cho2021vlt5} and Qwen2.5-VL-7B~\cite{bai2025qwen25vltechnicalreport}, which can process both text and visual signals. 

Our proposed test-time reranking (TTR) mechanism is specifically designed for decoder-based generative retrievers that support constrained decoding.
Accordingly, we evaluate its effectiveness on encoder-decoder models (T5, VLT5) and decoder-only models (Qwen2.5, Qwen2.5-VL), each under two configurations: with and without the TTR enabled. 

\subsection{Experimental Settings}
All baseline methods are incorporated into our proposed three-stage pipeline (see Figure~\ref{fig:framework_overview}), serving as the generative retriever component.
Each dataset is partitioned into training, validation, and testing splits at the dialogue level (see Table~\ref{tab:dataset_statistics}).
To ensure a fair comparison, all methods share the same training split and are trained using formats specific to their architectures: decoder-only models on full dialogues and encoder-decoder models on extracted user–system interaction pairs. Further training configurations are detailed in later sections.
Model selection is based on validation performance, and final results are reported on the test set. 

Retrieval effectiveness is evaluated at each conversational turn using standard ranking metrics: nDCG and MRR. 
Generated top-$b$ semantic IDs per product are set to 2, 2, and 5 for $\text{MFR}_{\text{crt}}$, $\text{MMD}_{\text{flt}}$, and $\text{MUSE}$, respectively.
Table~\ref{tab:main_res} reports the average scores for the final turns; the average scores across all turns are investigated in later sections. 
Note that each dialogue in our datasets corresponds to a single target item, differing from conventional single-turn text retrieval datasets such as \text{MS} \text{MARCO}~\cite{DBLP:conf/nips/NguyenRSGTMD16}, where a query may have multiple relevant answers. 
Following prior work~\cite{zhao-etal-2024-unifashion}, we construct a retrieval set of 100 candidates per query during testing, ensuring the inclusion of the target product.
This setup allows us to evaluate the effectiveness of our proposed test-time reranking mechanism in identifying and ranking the target product.

\subsubsection{Inferred Query Generation}
We follow the procedure outlined in the Methodology Section and employ GPT-4o-mini to generate inferred queries $\widehat{q_t}^{\rm text}$ for each turn $t$ in the training dialogues. 
Each inferred query is then paired with its corresponding image component (e.g., $q_t^{\rm img}$) to construct an inferred multimodal dialogue, which serves as the training data for the retriever.

\subsubsection{Retriever Training}
For decoder-only models, we employ the {LLaMA-Factory}~\cite{zheng2024llamafactory} toolkit to fine-tune the models using training dialogues in which user queries have been replaced with their inferred versions.
Each dialogue is serialized into a  {ShareGPT}-style token sequence for next-token prediction.
Training follows a standard autoregressive objective, with the loss computed only on tokens corresponding to the assistant's responses (i.e., $p_t$). 
If supported, visual inputs are encoded into visual tokens via a vision encoder and incorporated into the training sequence.

In contrast, encoder-decoder models are trained on pairwise data extracted from each user–system turn in the training dialogues. 
Given a dialogue $d$, we construct training pairs of the form $\{ (\hat{q}_1^{\scriptscriptstyle \mathcal{M}}, p_1),\; (\hat{q}_2^{\scriptscriptstyle \mathcal{M}}, p_2),\; \ldots,\; (\hat{q}_{|d|}^{\scriptscriptstyle \mathcal{M}}, p_{|d|}) \}$.
Like the decoder-only setup, training follows an autoregressive objective: given a inferred user query $\hat{q}_t^{\scriptscriptstyle \mathcal{M}}$, the model is trained to generate the associated product-side text, which may include the image caption, the product description, or both.
If supported, visual features are extracted using a vision encoder and integrated alongside the textual input.

\subsubsection{Test-time Reranking}
Following Equation~(\ref{eq:test-time product reranking}), we update the retrieval scores for models compatible with our proposed test-time reranking, namely, T5, VLT5, Qwen, and Qwen-VL. 
During the generative retrieval phase at inference time, beam sizes are determined per dataset and then shared across all models for fair comparisons.
We employ GPT-4o-mini as the evaluator model.

\subsubsection{FM-index for Inference}
During inference, we pre-encode the product corpus using an {FM-index}, a data structure chosen for its superior efficiency in constrained decoding over large corpora, outperforming alternatives like MarisaTrie~\cite{yata2011prefix}.
The FM-index is integrated into a {logits processor} that restricts the generative retriever's output to valid substrings corresponding to semantic IDs (SIDs) in the corpus~\cite{892127}.
To support this decoding constraint, we extended the widely-used {LLaMA-Factory}~\cite{zheng2024llamafactory} codebase by exposing additional functionalities for output constraint control through API---a previously unavailable feature. 
This extension enables compatibility with up-to-date MLLMs, including LLaVA~\cite{liu2023llava} and InternVL~\cite{chen2024internvl}. 
We will release our implementation upon paper acceptance.

\begin{table*}[!t]
\centering
\scalebox{0.92}{
\setlength{\tabcolsep}{5pt}
\begin{tabular}{cp{0.8cm}c*{8}{S[table-format=2.2,text-rm=\normalfont,detect-all=true]}}
\toprule
\multirow{2}{*}{{}} & \multirow{2}{*}{{\textbf{Dataset}}} & \multirow{2}{*}{{\textbf{~TTR}}} & \multicolumn{4}{c}{\textbf{Unimodal Model}} & \multicolumn{4}{c}{\textbf{Multimodal Model}} \\
\cmidrule(lr){4-7} \cmidrule(lr){8-11} & &
& {\textbf{~~~MRR~~~}} & {\textbf{nDCG@1}} & {\textbf{nDCG@5}} & {\textbf{nDCG@10}} & {\textbf{~~~MRR~~~}} & {\textbf{nDCG@1}} & {\textbf{nDCG@5}} & {\textbf{nDCG@10}} \\
\midrule
\multirow{12}{*}{{\rotatebox[origin=l]{90}{\textbf{\hspace{1.1cm} Decoder-only}}}} & \multicolumn{2}{c}{} & \multicolumn{4}{c}{Qwen2.5-7B} & \multicolumn{4}{c}{Qwen2.5-VL-7B} \\ 
\cmidrule(lr){2-3} \cmidrule(lr){4-7} \cmidrule(lr){8-11}
& \multirow{2}{*}{\texttt{$\text{MFR}_{\text{crt}}$}} & \xmark & 4.90 & 0.00 & 2.50 & 5.73 & 11.19 & 5.00 & 10.31 & 11.89 \\
~ & ~ & \cmark & \bfseries 25.11\,\normalfont\tiny{({\color{brightgreen}$\uparrow$}+ 20.2)} & \bfseries 15.00\,\normalfont\tiny{({\color{brightgreen}$\uparrow$}+ 15.0)} & \bfseries 23.24\,\normalfont\tiny{({\color{brightgreen}$\uparrow$}+ 20.7)} & \bfseries 27.86\,\normalfont\tiny{({\color{brightgreen}$\uparrow$}+ 22.1)} & \bfseries 23.11\,\normalfont\tiny{({\color{brightgreen}$\uparrow$}+ 11.9)} & \bfseries 15.00\,\normalfont\tiny{({\color{brightgreen}$\uparrow$}+ 10.0)} & \bfseries 22.50\,\normalfont\tiny{({\color{brightgreen}$\uparrow$}+ 12.2)} & \bfseries 23.95\,\normalfont\tiny{({\color{brightgreen}$\uparrow$}+ 12.1)} \\
\cmidrule(lr){2-3} \cmidrule(lr){4-7} \cmidrule(lr){8-11}
& \multirow{2}{*}{\texttt{$\text{MMD}_{\text{flt}}$}} & \xmark & 11.97 & 4.00 & 10.36 & 13.79 & 17.32 & 9.00 & 15.25 & 19.12 \\
~ & ~ & \cmark & \bfseries 36.33\,\normalfont\tiny{({\color{brightgreen}$\uparrow$}+ 24.4)} & \bfseries 23.00\,\normalfont\tiny{({\color{brightgreen}$\uparrow$}+ 19.0)} & \bfseries 37.76\,\normalfont\tiny{({\color{brightgreen}$\uparrow$}+ 27.4)} & \bfseries 41.70\,\normalfont\tiny{({\color{brightgreen}$\uparrow$}+ 27.9)} & \bfseries 38.59\,\normalfont\tiny{({\color{brightgreen}$\uparrow$}+ 21.3)} & \bfseries 25.00\,\normalfont\tiny{({\color{brightgreen}$\uparrow$}+ 16.0)} & \bfseries 39.61\,\normalfont\tiny{({\color{brightgreen}$\uparrow$}+ 24.4)} & \bfseries 44.31\,\normalfont\tiny{({\color{brightgreen}$\uparrow$}+ 25.2)} \\
\cmidrule(lr){2-3} \cmidrule(lr){4-7} \cmidrule(lr){8-11}
& \multirow{2}{*}{\texttt{MUSE}} & \xmark & 5.05 & 0.00 & 3.84 & 4.47 & 6.83 & 2.00 & 4.91 & 5.94 \\
~ & ~ & \cmark & \bfseries 12.23\,\normalfont\tiny{({\color{brightgreen}$\uparrow$}+ 7.2)} & \bfseries 4.00\,\normalfont\tiny{({\color{brightgreen}$\uparrow$}+ 4.0)} & \bfseries 11.19\,\normalfont\tiny{({\color{brightgreen}$\uparrow$}+ 7.4)} & \bfseries 13.81\,\normalfont\tiny{({\color{brightgreen}$\uparrow$}+ 9.3)} & \bfseries 11.45\,\normalfont\tiny{({\color{brightgreen}$\uparrow$}+ 4.6)} & \bfseries 5.00\,\normalfont\tiny{({\color{brightgreen}$\uparrow$}+ 3.0)} & \bfseries 10.61\,\normalfont\tiny{({\color{brightgreen}$\uparrow$}+ 5.7)} & \bfseries 11.52\,\normalfont\tiny{({\color{brightgreen}$\uparrow$}+ 5.6)} \\
\midrule
\multirow{10}{*}{{\rotatebox[origin=l]{90}{\textbf{ \hspace{0.2cm} Encoder-decoder}}}} & \multicolumn{2}{c}{} & \multicolumn{4}{c}{T5-Large} & \multicolumn{4}{c}{VLT5-Large} \\ 
\cmidrule(lr){2-3} \cmidrule(lr){4-7} \cmidrule(lr){8-11}
& \multirow{2}{*}{\texttt{$\text{MFR}_{\text{crt}}$}} & \xmark & 5.13 & 1.00 & 2.63 & 5.52 & 5.54 & 1.00 & 3.79 & 5.39 \\
~ & ~ & \cmark & \bfseries 10.58\,\normalfont\tiny{({\color{brightgreen}$\uparrow$}+ 5.5)} & \bfseries 3.00\,\normalfont\tiny{({\color{brightgreen}$\uparrow$}+ 2.0)} & \bfseries 9.36\,\normalfont\tiny{({\color{brightgreen}$\uparrow$}+ 6.7)} & \bfseries 12.86\,\normalfont\tiny{({\color{brightgreen}$\uparrow$}+ 7.3)} & \bfseries 11.38\,\normalfont\tiny{({\color{brightgreen}$\uparrow$}+ 5.8)} & \bfseries 6.00\,\normalfont\tiny{({\color{brightgreen}$\uparrow$}+ 5.0)} & \bfseries 8.52\,\normalfont\tiny{({\color{brightgreen}$\uparrow$}+ 4.7)} & \bfseries 12.35\,\normalfont\tiny{({\color{brightgreen}$\uparrow$}+ 6.9)} \\
\cmidrule(lr){2-3} \cmidrule(lr){4-7} \cmidrule(lr){8-11}
& \multirow{2}{*}{\texttt{$\text{MMD}_{\text{flt}}$}} & \xmark & 12.54 & 4.00 & 11.29 & 14.19 & 13.73 & 5.00 & 12.65 & 16.63 \\
~ & ~ & \cmark & \bfseries 34.70\,\normalfont\tiny{({\color{brightgreen}$\uparrow$}+ 22.2)} & \bfseries 22.00\,\normalfont\tiny{({\color{brightgreen}$\uparrow$}+ 18.0)} & \bfseries 35.63\,\normalfont\tiny{({\color{brightgreen}$\uparrow$}+ 24.3)} & \bfseries 40.13\,\normalfont\tiny{({\color{brightgreen}$\uparrow$}+ 25.9)} & \bfseries 37.89\,\normalfont\tiny{({\color{brightgreen}$\uparrow$}+ 24.2)} & \bfseries 24.00\,\normalfont\tiny{({\color{brightgreen}$\uparrow$}+ 19.0)} & \bfseries 38.27\,\normalfont\tiny{({\color{brightgreen}$\uparrow$}+ 25.6)} & \bfseries 43.48\,\normalfont\tiny{({\color{brightgreen}$\uparrow$}+ 26.9)} \\
\cmidrule(lr){2-3} \cmidrule(lr){4-7} \cmidrule(lr){8-11}
& \multirow{2}{*}{\texttt{MUSE}} & \xmark & 8.92 & 2.00 & 7.56 & 8.54 & 6.57 & 2.00 & 3.64 & 6.57 \\
~ & ~ & \cmark & \bfseries 20.10\,\normalfont\tiny{({\color{brightgreen}$\uparrow$}+ 11.2)} & \bfseries 11.00\,\normalfont\tiny{({\color{brightgreen}$\uparrow$}+ 9.0)} & \bfseries 17.59\,\normalfont\tiny{({\color{brightgreen}$\uparrow$}+ 10.0)} & \bfseries 23.06\,\normalfont\tiny{({\color{brightgreen}$\uparrow$}+ 14.5)} & \bfseries 22.18\,\normalfont\tiny{({\color{brightgreen}$\uparrow$}+ 15.6)} & \bfseries 9.00\,\normalfont\tiny{({\color{brightgreen}$\uparrow$}+ 7.0)} & \bfseries 22.61\,\normalfont\tiny{({\color{brightgreen}$\uparrow$}+ 19.0)} & \bfseries 27.50\,\normalfont\tiny{({\color{brightgreen}$\uparrow$}+ 20.9)} \\
\bottomrule
\end{tabular}
}
\caption{Evaluation results on the final-turn queries of testing dialogues across three datasets, comparing two model architectures (decoder-only and encoder–decoder) and two query modalities (unimodal and multimodal).
We report the average retrieval scores for each model with and without \textbf{test-time reranking (TTR)}.
TTR results are shown in \textbf{bold} when they outperform the corresponding non-TTR baseline, with the performance gain indicated in parentheses as {\tiny $({\color{brightgreen}\uparrow}\mathrm{+xx.x})$} following the bolded value.
}
\label{tab:main_res}
\end{table*}

\subsection{Experimental Results}
\subsubsection{Effectiveness of Test-time Reranking}
Table~\ref{tab:main_res} reports retrieval performance on the final-turn user queries across three datasets.
The results demonstrate that our proposed test-time reranking (TTR) consistently improves key retrieval metrics, including MRR and nDCG. For instance, with the Qwen2.5-VL model, TTR yields substantial gains across all datasets: on $\text{MFR}_{\text{crt}}$, MRR improves by 12 points (a 107\% relative increase, from 11.19 to 23.11; on $\text{MMD}_{\text{flt}}$, by 21 points (a relative 123\% increase, from 17.32 to 38.59); and on MUSE, by 5 points (a relative 68\% increase, from 6.83 to 11.45). 
Similarly, nDCG@1 scores increase by 10 points on $\text{MFR}_{\text{crt}}$ (from 5 to 15), 16 points on $\text{MMD}_{\text{flt}}$ (from 9 to 25), and 3 points on MUSE (from 2 to 5).
Notablely, TTR proves effective across both decoder-only and encoder–decoder model architectures: all four models---Qwen2.5, Qwen2.5-VL, T5, and VLT5---exhibit consistent performance gains at inference time when TTR is applied.

\subsubsection{Effectiveness of Multimodal Query}
We also compare the effectiveness of unimodal and multimodal query settings. 
As shown in Table~\ref{tab:main_res}, models that support multimodal queries (i.e., text accompanied by a reference image) generally outperform their text-only counterparts. 
After applying TTR, VLT5 achieves an average MRR gain of 2 points over T5 across all three datasets.
Similarly, even without TTR, Qwen2.5-VL outperforms its unimodal variant Qwen2.5 on all benchmarks.
Notably, both Qwen2.5-VL and Qwen2.5 exhibit comparable gains in retrieval performance after TTR is applied.
These findings confirm that incorporating visual context into user queries enhances retrieval effectiveness.

\subsubsection{Checkpoint Selection}
We further examine the impact of TTR on the generative retriever across different training checkpoints. 
Using the $\text{MMD}_{\text{flt}}$ dataset---the largest in our study---we evaluate four Qwen2.5-VL checkpoints spanning early to late training stages, applying TTR with fixed hyperparameters and reporting average MRR scores over test set final turns.
As shown in Figure~\ref{fig:checkpoint_analysis}, our analysis reveals that while retrieval performance degrades at later checkpoints due to overfitting, TTR effectively mitigates this issue by reassessing the semantic alignment between retrieved products and the user's evolving intent. 
The performance gains are especially pronounced at later stages, where the model exhibits a stronger bias toward hard negatives. 
These results demonstrate that applying test-time scaling to selected checkpoints enhances retrieval robustness and overall effectiveness.

\subsubsection{Retrieval Performance across Dialogue Turns}

\begin{figure}[!htb]
\begin{minipage}[b]{0.52\textwidth}
    \centering
    \includegraphics[width=\textwidth]{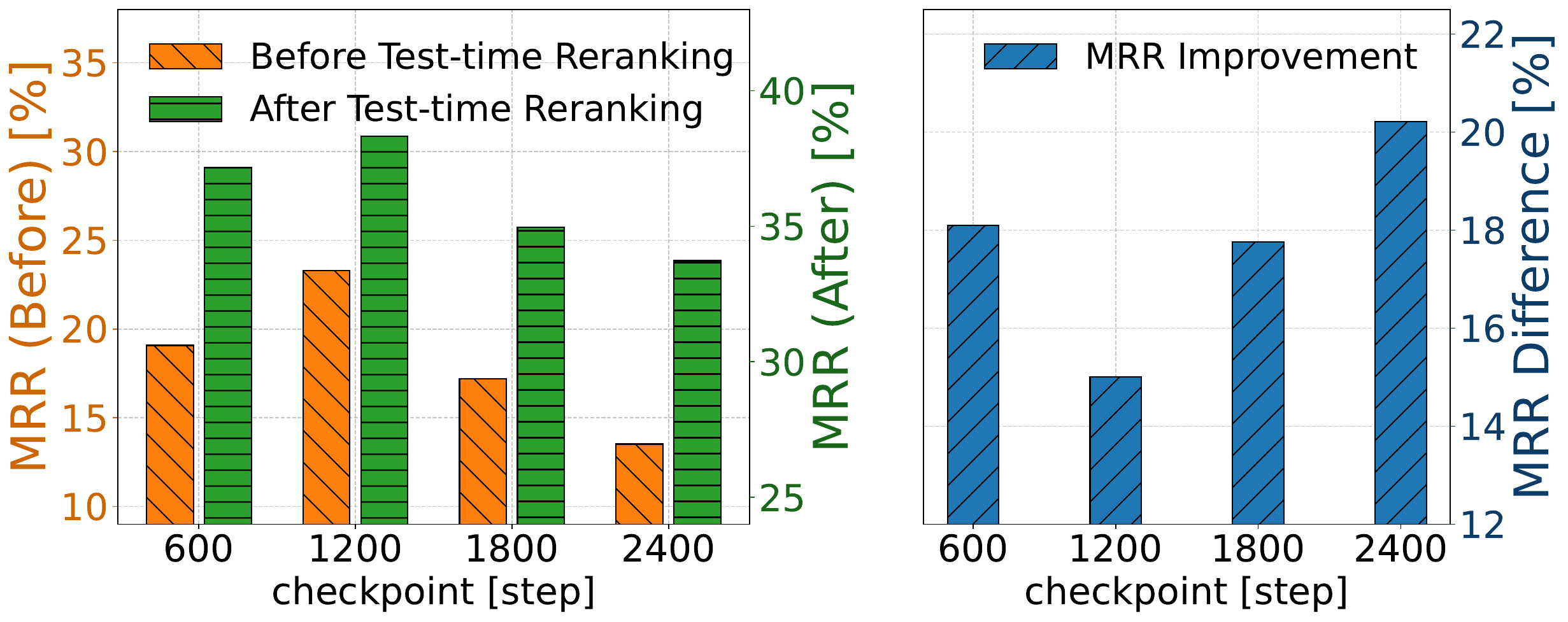}
    \caption{TTR performance across different checkpoints.}
    \label{fig:checkpoint_analysis}
\end{minipage}
\hfill
\begin{minipage}[b]{0.45\textwidth}
\centering
\scalebox{0.8}{
\setlength{\tabcolsep}{2.5pt}
\begin{tabular}{ccccccccc}
\toprule
\multirow{2}{*}{{\textbf{Dataset}}} & \multirow{2}{*}{\textbf{TTR}} & \multicolumn{6}{c}{\textbf{Turn}} \\
\cmidrule(lr){3-8} &
& \texttt{1} & \texttt{2} & \texttt{3} & \texttt{4} & \texttt{5} & \texttt{6} \\
\midrule
\multirow{2}{*}{\texttt{$\text{MFR}_{\text{crt}}$}} & \xmark & 3.43 & 3.32 & 3.52 & 8.21 & - & - \\
~ & \cmark & \textbf{3.94} & \textbf{13.29} & \textbf{13.00} & \textbf{24.93} & - & - \\
\cmidrule(lr){1-2} \cmidrule(lr){3-8}
\multirow{2}{*}{\texttt{$\text{MMD}_{\text{flt}}$}} & \xmark & 11.39 & 6.37 & 9.27 & 9.37 & 11.69 & 19.58 \\
~ & \cmark & \textbf{29.83} & \textbf{28.26} & \textbf{22.31} & \textbf{25.16} & \textbf{37.41} & \textbf{39.33} \\
\cmidrule(lr){1-2} \cmidrule(lr){3-8}
\multirow{2}{*}{\texttt{MUSE}} & \xmark & 1.37 & 5.03 & 5.26 & 6.92 & - & - \\
~ & \cmark & \textbf{1.96} & \textbf{8.82} & \textbf{10.46} & \textbf{11.67} & - & - \\

\bottomrule

\end{tabular}
}
\vspace{0.2cm}
\captionof{table}{Turn-specific average MRR scores.}
\label{tab:retrieval_performance_at_each_turn}
\end{minipage}
\end{figure}

To assess retrieval accuracy over a conversation, we follow prior work~\cite{10.1145/3589334.3645483} and report the average MRR at each dialogue turn across the three datasets using Qwen2.5-VL.
To ensure comparability across turns, we select test dialogues of uniform length: 11 dialogues with 4 turns from $\text{MFR}_{\text{crt}}$, 25 with 6 turns from $\text{MMD}_{\text{flt}}$, and 60 with 4 turns from MUSE.
As shown in Table~\ref{tab:retrieval_performance_at_each_turn}, the proposed TTR consistently improves retrieval performance, with particularly notable gains in the later turns. 
We also observe a general upward trend in MRR as conversations progresses, indicating that accumulating dialogue history plays a critical role in enhancing retrieval accuracy over multiple turns.

\section{Conclusions}
In this paper, we extend test-time scaling to the domain of multimodal conversational product retrieval by proposing a novel framework.
Our approach integrates a semantic ID-based generative retriever with a multimodal conversational retrieval pipeline and introduces a test-time reranking (TTR) mechanism that verifies the retriever's prediction---such as product attributes and item relevance---against the user's evolving intent during inference.
This mechanism significantly enhances retrieval accuracy and recommendation quality in multi-turn product search scenarios.

We evaluate two model architectures (i.e., decoder-only and encoder-decoder) across three benchmark datasets, showing that TTR consistently enhances retrieval performance. Additional analyses, including checkpoint sensitivity and turn-level performance, further validate the effectiveness and robustness of our approach.
Overall, our findings show that TTR yields significant performance gains in multimodal conversational product search, with minimal inference-time overhead.
To support further research, we release curated multi-turn multimodal retrieval datasets, augmented with our inferred queries and annotated target products. These resources include improved versions of existing benchmarks and a synthetic dataset generated using a computationally intensive method established in the field.

\bibliographystyle{unsrt}  
\bibliography{references}  

\end{document}